\documentclass[
nopreprint,
]{jasatex}

\usepackage{graphicx}
\usepackage{dcolumn}
\usepackage{amsmath,amsfonts}
\usepackage{bm}
\usepackage{multirow}

\begin{document}

\preprint{AIP/123-QED}

\title[Acoustic Radiation Losses of Micro Resonators]{Contribution of Acoustic Losses in the Quality Factor of a Micromechanical Resonator}

\author{Santhosh D. Vishwakarma}
 \email{santdmec@mecheng.iisc.ernet.in}
 \altaffiliation[Also at ]{Department of Mechanical Engineering, Indian Institute of Science, Bengaluru 560012, Karnataka, India}
\affiliation{Centre for Nano Science and Engineering, Indian Institute of Science, Bengaluru 560012, India}
\author{A. K. Pandey}
\affiliation{Department of Mechanical Engineering, Indian Institute of Technology, Hyderabad 502205, India}%
\author{J. M. Parpia}
 \affiliation{Physics Department, Clark Hall, Cornell University, Ithaca, New York 14853-2501, USA}%
\author{R. Pratap}%
\affiliation{%
Department of Mechanical Engineering, Indian Institute of Science, Bengaluru 560012, India}
\date{\today}
\begin{abstract}
A semi-analytical study of the acoustic radiation losses associated with various transverse vibration modes of a micromechanical (MEMS) annular resonator is presented. The quality factor, Q, of such resonators is of interest in many applications and depends on structural geometry, interaction with the external environment, and the encapsulation method. Resonators with at least one surface exposed to air can display losses through acoustic radiation even at $\mu$m dimensions. Published experimental results suggest the dominance of acoustic losses in the Q of a MEMS drum resonator. In this study, a well established mathematical techniques to analytically model resonator vibration modes and fluid-structure interaction are used, and a semi-analytical procedure for computing Q due to acoustic radiation losses, Qac, in any vibrational mode outlined. Present technique includes calculation of the exact mode shape and its utilization in computing Qac. The dependence of Qac on the first 15 mode shapes is computed. Results are compared for the lowest 2 modes of a solid circular resonator using exact mode shapes to those of Lamb's approximate mode shapes. Comparison to published experimental results validates the predictive utility of the technique, especially for higher modes where acoustic radiation seems to be the dominant constituent of Q.
\end{abstract}
\pacs{4340Rj, 4340Dx, 4340Ey, 4338Gy}
\maketitle

\section{\label{sec:level1}Introduction:\protect}

Mechanical resonators can have significant advantages such as small size, low power
operation\cite{Clark} and the ability to be integrated into CMOS structures\cite{Maxim} yielding high
quality factor (Q) devices\cite{Vivek} for use as RF filters\cite{Chandrahalim} and for mass sensing\cite{Yang}. For chemical
sensing, the route that uses changes in stress to monitor the presence of trace
vapors from the ambient air is particularly attractive\cite{Wachter,Datskos,Darrenstress}. In this sensing method, a polymer coating applied to one side of the mechanical resonator changes its
morphology, swelling or contracting and thus altering the stress in the underlying
silicon structure, which in turn produces a shift in the resonant frequency.
However, the fact that the vibrating mechanical structure must be immersed in
air so as to enable the sensing mechanism has consequences, since the presence of
the air leads to mechanical energy loss to the medium. The loss increased
dissipation Q$^{-1}$, is manifested as a broadening of the resonant peak,
leading to a degradation of the ability to resolve frequency shift and consequently
reduction in the sensitivity to analyte concentrations.

It has long been recognized that there are two main sources of energy loss of such
micron scale resonators. The first, termed squeeze film losses are due to the
trapped film of air under a vibrating resonator\cite{MinhangBaoreview,MinhangBao}. The second loss mechanism is where
the vibrating structure couples acoustically to the air. This paper quantifies the
loss due to acoustic radiation for a conceptually simple readily fabricated
structure: an annular plate fixed at its edges. We show that the
acoustic losses can be reduced by operating the structure at higher harmonics, where
the phase difference between adjacent moving segments at various vibrating modes of the resonator leads to a reduction in radiated energy.

A circular plate vibrating in contact with surrounding fluid, is of significant interest in wide range of systems such as the piston movement in contact with fluid in a closed cylinder \cite{Kinsler}, ocean structure moving underwater \cite{Lamb}, movement of magnetic disk drive in contact with air \cite{Rajendra}, vibration of nano circular drum resonator in contact with surrounding fluid in case of pressure sensors \cite{Darren}, microfluidic device with circular membrane \cite{Pratap}, etc. There are two main effects associated with such systems due to the coupled effect of fluid-structure interaction. In the first, the ``added mass effect'' reduces the effective resonance frequency. In the second, the acoustic radiation affects the quality factor of the vibrating structure. This fluid-structure problem is analyzed numerically using the finite element method or boundary element method for complex domains and analytically for simple domains. In this paper, we present an analytical procedure to analyze the acoustic radiation losses in higher modes of the vibrating annular plate with fixed outer boundaries.

Lord Rayleigh \cite{Rayleigh2} was probably the first to study analytically the effect of the mass loading of the surrounding fluid on a vibrating rigid disk in contact with the fluid and suggest the idea of ``added mass''. This classic problem was then studied by Lamb \cite{Lamb} to investigate the added mass induced change in the first two natural frequencies of a circular plate, fixed along its outer edge and vibrating in contact with a fluid. In addition to the frequency change, he also found the acoustic radiation losses corresponding to those two modes.  These results have been validated experimentally by subsequent studies. It is, however, important to note that Lamb's results are based on approximate mode shapes. Revisiting this problem, Amabili \emph{et al}. \cite{Amabilisolid,Amabiliannular} used the Hankel transform to analyze the added mass effect on the frequency of the structure vibrating in its fundamental and higher modes. However, their study was limited to analysis of the added mass effect. Such fluid-structure interaction effects have recently been analyzed in the context of Micro and Nano Electro Mechanical Systems (MEMS/NEMS) structures. Unlike in the case of macroscale problems where the surrounding fluid is assumed to be incompressible and inviscid, the  experimental validation of Lamb's theory was done examined for a microsensor working in the presence of a viscous fluid (water-glycerol mixture) \cite{Ayela}. Experimental results were found to be in good agreement with Lamb's predictions for a less viscous fluid mixture ($<$ 10 cP) but differed for fluid mixtures with higher viscosity. The difference was attributed to the viscosity contribution to the added mass effect qualitatively \cite{Ayela} as well as quantitatively \cite{Kozlovsky}. Considering the first approximate mode shape of a circular plate vibrating in contact with the surrounding fluid, Kozlovsky \cite{Kozlovsky} analyzed the effect of viscosity on the natural frequency as well as the quality factor. Recently, Olfatnia \emph{et al}. \cite{Olfatnia} have compared  theoretical and experimental results for analyzing the effect of viscosity of the surrounding fluid on the frequency as well as the quality factor of a circular diaphragm vibrating in its first mode.

Our central interest in the present study is to find acoustic losses and the associated quality factor in various modes of vibration of an annular micromechanical resonator (a MEMS plate) clamped at its outer edge in order to assess the suitability of higher modes to high-Q applications. The annularity of the resonator results from the requirement of an etch hole typically used in the micromachining technique to create a cavity underneath the resonator by etching away the supporting oxide material starting from the etch hole. Although the etch holes can be many and spread over the plate, a central etch hole serves the purpose and creates the simplest resonator structure \cite{Darren}. The annular resonator is also particularly amenable to analytical treatment for studying acoustic losses and building mathematical models for predicting the Q-factor.

In order to analyze acoustic radiation losses associated with various modes of vibration of the annular plate, we first derive the exact mode shapes of the structure ignoring any effect of the surrounding fluid (air) on the mode shape. We use these mode shapes to study the effect of the surrounding fluid on the associated natural frequencies and the Q-factor. Since the surrounding fluid is air, the effect of ``added mass'' on the frequencies of the structure is negligible. The effect of the surrounding air on the Q-factor, however, is significant because of the acoustic radiation losses which is the subject matter of this paper. We extend the analytical approach proposed by Amabili \emph{et al}. \cite{Amabilisolid,Amabiliannular} to higher modes, establish an analytical calculation procedure involving symbolic algebra, find the acoustic radiation losses, and compare our results, first, with Lamb's results for the first two modes, and then some published experimental results\cite{Darren} for the higher modes of the resonator.

\section{\label{sec:level1}Mathematical Modeling \protect}
\begin{figure}[h]
\includegraphics[width=8cm,clip]{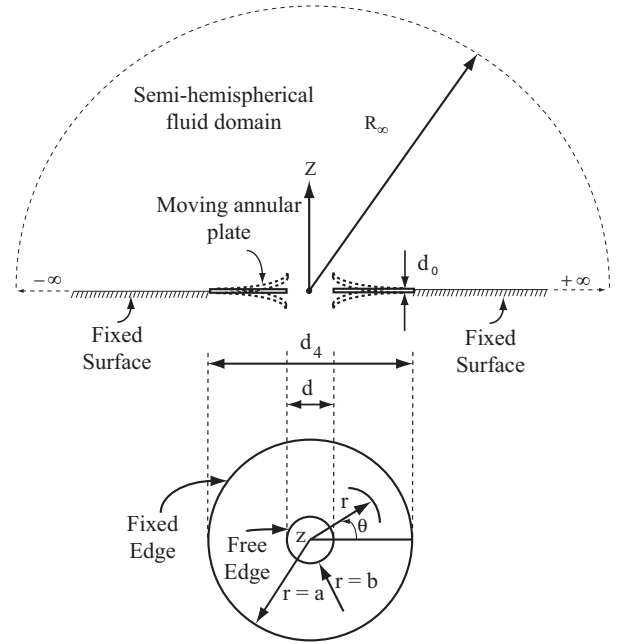}
\caption{Schematic of semi-hemispherical fluid domain  above the vibrating annular plate fixed at its outer edge}
\label{Acous}
\end{figure}

In this section, we present the mathematical background for computing the acoustic radiation losses due to a vibrating annular plate of free inner edge and fixed outer edge. The annular plate, with the inner radius $a$ and the outer radius $b$ in contact with the hemispherical surrounding fluid on its upper surface as shown in Fig.~\ref{Acous}.

First, we find the expression for the exact mode shapes of the vibrating annular plate in vacuum. Subsequently, using the same mode shapes, we present the procedure for computing the acoustic radiation losses.

\subsection{\label{sec:Mode}Modal Analysis of Annular Plate }

The vibration of annular plates with fixed outer boundary is widely analyzed for different applications. Under the assumptions of a thin plate made of isotropic, homogeneous, and linearly elastic material, the equation governing the transverse deflection $W(r,\theta,t)$ of the plate for undamped, free vibration can be written in polar co-ordinates as, \cite{Vogel, Raju,JRao}
\begin{equation}
\nabla^{4} W + \left(\frac{\rho_{s}d_{0}}{D}\right) \frac{\partial^{2}W}{\partial t^{2}} = 0,
\label{3}
\end{equation}
where, the Laplacian operator $\nabla^{2}=\frac{\partial^{2}}{\partial r^{2}}+\frac{1}{r}\frac{\partial}{\partial r} + \frac{1}{r^{2}} \frac{\partial^{2}}{\partial \theta^{2}}$ and $D = \frac{Ed_{0}^3}{12(1-\nu^2)}$ is flexural rigidity of the plate, $\rho_{s}$ is the density and $d_{0}$ is the uniform thickness of the plate. Assuming that the plate vibrates in a normal mode, $W$ can be expressed as,
\begin{equation}
 W  = w(r,\theta)e^{i \omega t}.
\label{4}
\end{equation}
Substituting eqn.~(\ref{4}) in eqn.~(\ref{3}) and then rearranging the resulting equation, we get the modal equation as,
\begin{equation}
 \nabla^{4} w(r,\theta)-\beta^{4} w(r,\theta) = 0,
\label{6}
\end{equation}
where, $\beta^{4} = \frac{\rho_{s}d_{0}}{D}\omega_{\rm{dry}}^{2}$ is the frequency
parameter. The modal solution $w(r,\theta)$ can be expressed in terms of Bessel functions as described in Appendix (\ref{A00}) and is given by, \cite{Watson}

\begin{multline}
w(r,\theta) = \chi(r)\psi(\theta) = [A_{mn}J_n(\beta r)+B_{mn}Y_{n}(\beta r)+\\
C_{mn}I_n(\beta r)+D_{mn}K_{n}(\beta r)]\psi(\theta), \label{18}
\end{multline}

where, $J_{n}$ is the Bessel function of the first kind, $Y_{n}$ is the Bessel function of the second kind, $I_{n}$ and $K_{n}$ are the modified Bessel's functions of the first and second kind, respectively.

For the annular plate with the fixed outer edge and free inner edge as shown in Fig.~\ref{Acous}, the following boundary conditions can be used. At the outer radius $r=a$, $w = \frac{\partial w}{\partial r} = 0$ and at the inner radius $r=b$ the bending moment \cite{Timoshenko} $M_{r}=-D\left[\frac{\partial^{2}w}{\partial r^{2}}+\nu(\frac{1}{r}\frac{\partial w}{\partial r}+\frac{1}{r^{2}}\frac{\partial^{2}w}{\partial\theta^{2}})\right] = 0$ and the shear force\cite{Timoshenko} $V_{r} =-D\left[\frac{\partial}{\partial r}\nabla^{2}w+\frac{1-\nu}{r^{2}}\frac{\partial^{2}}{\partial{\theta}^{2}}(\frac{\partial
w}{\partial r}-\frac{w}{r})\right]= 0$ . On applying these boundary conditions, we get the system of four linear and homogenous equations for the four constants $A_{mn}$, $B_{mn}$, $C_{mn}$ and $D_{mn}$ which we list as eqns.(\ref{19})-(\ref{22}) in Appendix \ref{A0}. For a non-trivial solution of these constants, the determinant of the coefficient matrix given by eqn.~(\ref{annu1}1), Appendix D, of the above equations is set to zero, which gives the required characteristic equation governing the frequency constant $\beta$. Since the characteristics equation given by eqn.~(\ref{annu1}1), Appendix D, is difficult to solve analytically, we employ a numerical technique based on the bisection method for finding roots in MATLAB to estimate $\beta$. The different solutions of $\beta$ are the natural frequencies, $f_{\rm{dry}}$, for different modes. For a given value of $\beta$, the corresponding constants $A_{mn}$, $B_{mn}$, $C_{mn}$, and $D_{mn}$ can be determined by solving the system of linear homogenous equations~(\ref{19})-(\ref{22}). Since the equations are homogenous, these constants can be expressed in terms of any one (say $D_{mn}$) of these constants. Therefore, expressing the constants $A_{mn}$, $B_{mn}$, $C_{mn}$ in terms of $D_{mn}$ and substituting them in eqn.~(\ref{18}), the mode shape can be written as $\frac{w(r,\theta)}{\psi(\theta)} = D_{mn}\tau_{mn}(\beta r)$. Normalizing the mode shape based on the normalization condition, \cite{Amabiliannular}
\begin{equation}
\int_{b/a}^{1} \left(\frac{w(\alpha,\theta)}{\psi(\theta)}\right)^{2}\alpha d{\alpha} = 1,
\label{38}
\end{equation}
where, $\alpha = {r}/{a}$, we get the expression of $D_{mn}$ as
\begin{equation}
D_{mn} = \frac{1}{\sqrt{{\int_\frac{b}{a}^1\{\tau_{mn}(\beta \alpha a)}\}^{2}\alpha d\alpha}}.
\label{42}
\end{equation}
For a particular resonant mode of vibration of the annular plate, the computed $D_{mn}$ from eqn.~(\ref{42}) can be used to determine other constants $A_{mn}$, $B_{mn}$ and $C_{mn}$ respectively.

For the limiting condition of the annular plate when the radius of the inner hole goes to zero, i.e., a solid plate, we find, following a similar procedure as described above, that $Y_{n}$ and $K_{n}$ tend to infinity as $r$ tends to zero \cite{Watson}. As $w$ is finite at the center of the plate, we must set $B_{mn}$ and $D_{mn}$ to be zero. The resulting deflection profile takes the form, \cite{JRao,Watson}
\begin{equation}
w(r,\theta) = [A_{mn}J_n(\beta r)+ C_{mn}I_n(\beta r)]\psi(\theta).
\label{32}
\end{equation}
Using the boundary conditions and following a similar procedure as described above, we get the following characteristic equation,
\begin{equation}
J_n(\beta a)I_{n+1}(\beta a)+I_n(\beta a)J_{n+1}(\beta a)=0.
\label{37}
\end{equation}
Again, solving the above characteristic equation for $\beta$ and using the normalization condition, we can obtain expression for values of $A_{mn}$ and $C_{mn}$.

Most of the fluid structure interaction problems\cite{Lamb} use a polynomial approximation for the mode shapes. However, such approximation introduces errors in estimating the damping from the fluid-structure interaction.
In the present analysis, we use exact mode shapes for estimating the acoustic radiation losses in different mode shapes.

\subsection{\label{sec:level3} Estimation of Acoustic Radiation Losses}
In the previous section, we found the natural frequencies and corresponding mode shapes of the annular plate vibrating in vacuum. When the plate vibrates in contact with the relatively denser surrounding fluid than air, the change in the mode shape of the plate is assumed to be negligible but there is a non-negligible reduction in the modal frequency. Such hypothesis has been used in many fluid-structure interaction problems \cite{KWAK,Amabilisolid,Amabiliannular}. However, we consider the corresponding changes in frequency in our formulation for estimating the acoustic radiation losses in the surrounding fluid.

As the annular plate vibrates, a disturbance is created in the fluid adjacent to the plate, which causes the wave motion to introduce the pressure fluctuations at all points in the fluid domain \cite{LAMBB}. Considering the surrounding fluid as irrotational and inviscid at constant ambient mean pressure $p_{0}$, temperature $T_{0}$ and density $\rho_{f}$, the pressure fluctuation above $p_{0}$ and the particle velocity can be found in terms of the velocity potential $\phi$ through the equation, \cite{Kinsler,LAMBB}
\begin{equation}
p=-\rho_{f} \frac{\partial \phi}{\partial t} ~ {\rm and}~ v=\nabla \phi. \label{pres_vel_potential}
\end{equation}
The governing equation for the velocity potential corresponding to small disturbances is given by, \cite{LAMBB,Kinsler,skudrzyk}
\begin{equation}
\nabla^{2}\phi-\frac{1}{c_{s}^{2}}\frac{\partial^2\phi}{\partial t^2} = 0,
\label{waveeq}
\end{equation}
where $c_{s}$ is the speed of acoustic waves in the fluid. The incompressibility condition $\nabla.v=0$ leads to the Laplace equation in terms of $\phi$, i.e., $\nabla^{2}{\phi}=0$. Therefore, it is sufficient to find the velocity potential field in order to analyze the propagation of waves in the fluid medium.

For the domain shown in Fig.~\ref{Acous}, the fixed outer support (i.e., for $r>a$) is assumed to be radially extended to infinity. For the hemispherical fluid domain enclosing the upper surface of the annular plate and the support, along with the Sommerfield boundary conditions, i.e., velocity and its gradient vanish as $r \rightarrow \infty$, the velocity potential can be obtained by solving the wave equation as mentioned above.

\begin{figure}[h]
\includegraphics{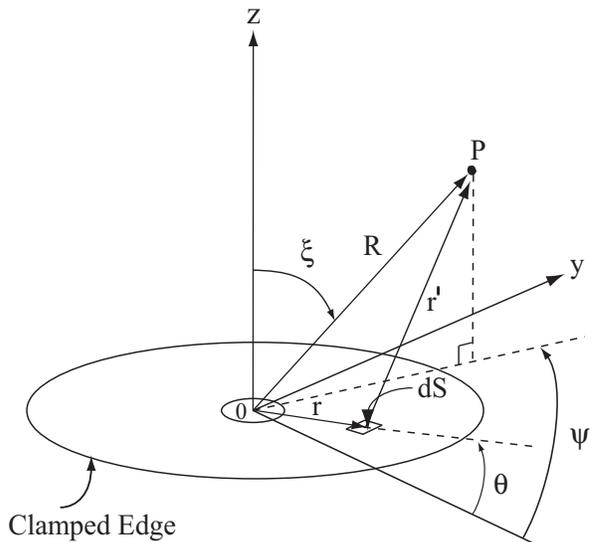}
\caption{Generic point $P$ in acoustic far field from the resonator}
\label{Acousticfield}
\end{figure}

\begin{figure}[htbp]
\includegraphics[width=19cm,height=20cm,clip]{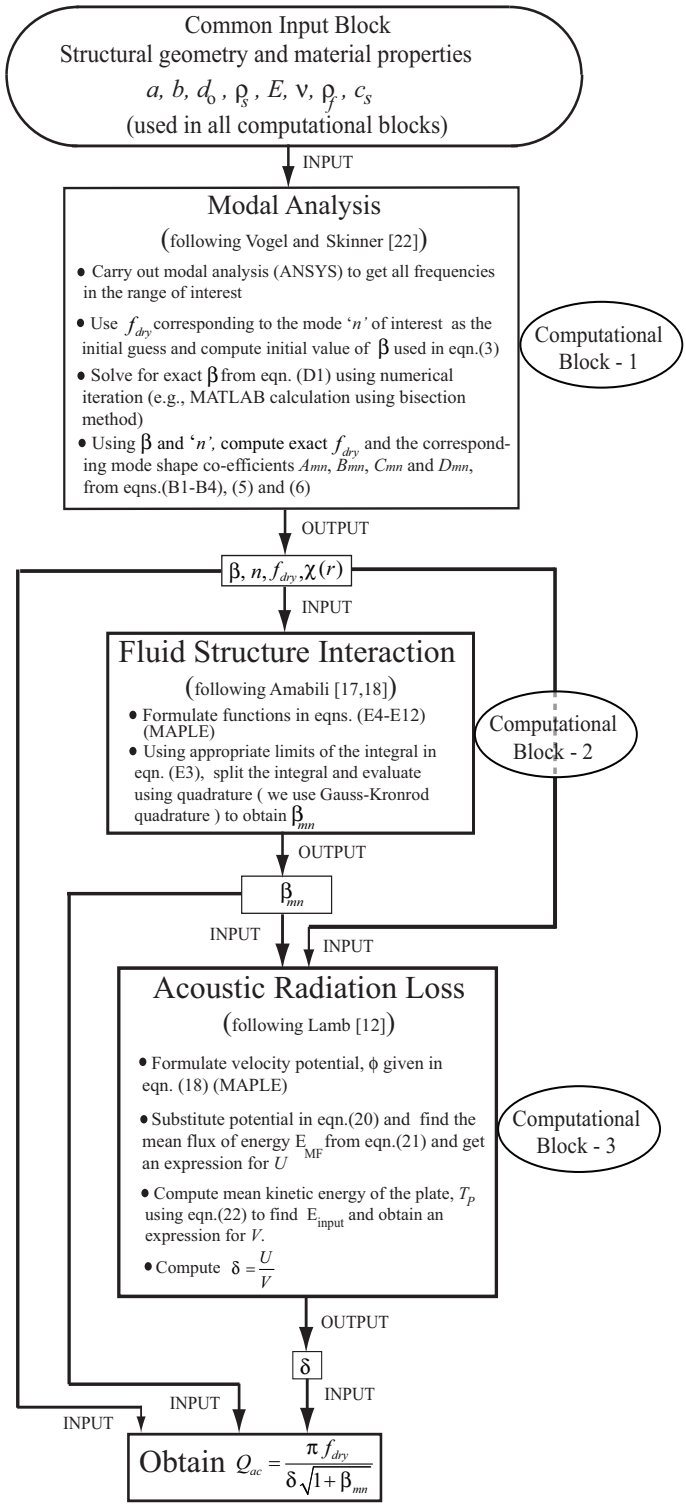}
\caption{Flow chart for computing acoustic radiation damping}
\label{flowchart}
\end{figure}

To find the velocity potential at a generic point $P$ due to an elementary source at $S$ as shown in Fig.~\ref{Acousticfield}, we follow the analysis given by Lamb\cite{Lamb}. Let the position co-ordinates of point $P$ be  $(R \sin{\xi} \cos{\psi}, R \sin{\xi} \sin{\psi} ,R\cos{\xi})$,
and that of $S$ on elemental surface area $dS$ in the plane
of the resonator be $S(r\cos{\theta}, r \sin{\theta}, 0)$. The distance $r^{\prime}$ of the
the point $P$ from the source position $S$ is given by
\begin{equation}
r^{\prime} = \left\{R^{2}-2R r \sin{\xi} \cos(\psi-\theta)+r^{2}\right\}^{\frac{1}{2}}.
\label{55}
\end{equation}
Since at the far field point $P$, $R \gg r$, the expression for $r^{\prime}$ can be approximated as,
\begin{equation}
r^{\prime} = \left\{R-r \sin{\xi} \cos(\psi-\theta)\right\}.
\label{56}
\end{equation}
Assuming that the normal component of the velocity of the fluid at the plate interface is same as that of the plate, the fluid velocity $v_{n}=\frac{\partial W}{\partial t}$ at point $S$ on the upper surface of the annular plate can be written in terms of the mode shape $w(r,\theta)$ defined in section \ref{sec:Mode}. When the energy loss in the form of acoustic radiation is accounted for (damped structural response), we need to modify eqn.~(\ref{4}) to accomodate a time varying amplitude that represents the transient decay. Assuming $W  = w(r,\theta)e^{i \omega t}$, we can find the normal velocity $v_{n}=\frac{\partial W}{\partial t}$ as,
\begin{equation}
\frac{\partial W}{\partial t}=i \omega~w(r,\theta) e^{i\omega t}
\label{159}
\end{equation}
But, $w(r,\theta) = \chi(r)\psi(\theta) = \chi(r) C \cos{n\theta}$, (see Appendix (A)). Therefore, $v_{n} = Ci\omega\chi(r)\cos{n\theta}e^{i\omega t}$. \\ Let $v_{n} = A\chi(r)\cos{n\theta}e^{i\omega t}$, where $A = iC\omega$.
We also know that because of acoustic radiation losses, the amplitude $A$ has to be time dependent, but its time dependence remains to be determined. So, at this point, we assume that,
\begin{equation}
v_{n} = A(t)\chi(r) \cos{n\theta} e^{i\omega t},
\label{59}
\end{equation}
where $\chi(r)=[A_{mn}J_n(\beta r)+B_{mn}Y_{n}(\beta r)+ C_{mn}I_n(\beta r)+D_{mn}K_{n}(\beta r)]$, $\omega$ is the frequency of the vibrating plate when in contact with the fluid. The velocity potential at a distance $R$ from the center of the resonator, or at a distance $r^\prime$ from the elementary surface $dS$ due to the disturbance at $S$ is given by, \cite{Rayleigh,Lamb}
\begin{equation}
\phi = \frac{1}{2\pi}\int_0^{2\pi}\int_b^a{\frac{e^{-ikr^{\prime}}}{r^{\prime}}\frac{\partial W}{\partial t}dS}.
\label{57}
\end{equation}
Using the approximation of $r^{\prime}$ from eqn.~(\ref{56}) and the expression of $\frac{\partial W}{\partial t}$ from eqn.~(\ref{59}), the velocity potential can be rewritten as,
\begin{multline}
\phi = \frac{A(t)e^{ik(c_{s}t-R)}}{2\pi
R}\int_b^{a}\int_0^{2\pi}{e^{ikr\rm{sin} \xi \cos(\psi-\theta)}}
\\ \times \cos{n\theta} \chi(r) r \rm{d}r {\rm{d}}\theta
\label{61}
\end{multline}
%
where $k=\frac{\omega}{c_{s}}$ is the wave number.
Using the following property of Bessel functions, \cite{skudrzyk,Rajendra}
%
%
\begin{equation}
\int_0^{2\pi}{e^{ikrsin \xi \cos(\psi-\theta)}} \cos{n\theta} {\rm{d}}\theta
= 2\pi i^{n} \cos{n\psi} J_{n}(kr \sin{\xi} ), \label{62}
\end{equation}
the velocity potential becomes,
%
\begin{equation}
\phi = \frac{A(t)i^{n}e^{ik(c_{s}t-R)}}{R} \cos{n\psi}
\times \int_b^{a}{J_{n}(k r \sin{\xi} ) \chi(r) r {\rm{d}}r}. \label{63}
\end{equation}
From this expression of velocity potential, one can calculate the pressure fluctuation and velocity using eqn.~(\ref{pres_vel_potential}).
Taking the real part of the velocity potential,
\begin{multline}
\phi_{r} = {\rm{Re}}\left(\frac{A(t)i^{n}e^{ik(c_{s}t-R)}}{R}\right)\cos{n\psi}\\
\times \int_b^{a}{J_{n}(kr \sin{\xi} ) \chi(r) r {\rm{d}}r}, \label{64}
\end{multline}

%
corresponding to the real part of the velocity $\frac{\partial W}{\partial t}=\chi(r) \cos{(n\theta)} \cos{(k c_{s}t)}$, the intensity of the acoustic wave at any point at distance $r^{\prime}$ can be written as the product of pressure and particle velocity at that point \cite{Kinsler}. The corresponding power radiated across a hemispherical surface of radius $R$, is obtained by integrating the intensity over the hemispherical fluid domain as
%
%
\begin{equation}
\frac{{\rm{d}} E_{\rm{flux}}}{{\rm{d}}t} =
\int_0^{{2\pi}}\int_0^{\frac{\pi}{2}}-\rho_{f}\left(\frac{\partial{\phi_{r}}}{\partial
t}\frac{\partial{\phi_{r}}}{\partial R}R^{2}\sin \xi \right){\rm{d}}\xi
{\rm{d}}\psi. \label{65}
\end{equation}
The mean flux of energy emitted in the form of sound waves is given by
\begin{equation}
E_{\rm{MF}} = \frac{\int_0^{\frac{2\pi}{\omega_{\rm{wet}}}}\frac{dE_{\rm{flux}}}{dt}dt}{\int_0^\frac{2\pi}{\omega_{\rm{wet}}}dt} = U A^{2}(t).
\label{66}
\end{equation}
 The expression for $U$ is obtained by substituting the velocity potential, $\phi_{r}$ in the power radiated eqn.(\ref{65}) and performing time averaging as in eqn.(\ref{66}). The parameter $U$ depends on fluid properties, geometric properties and frequency of oscillation of the resonator and is given by,
\begin{multline}
U = -\rho_{f}R
\frac{\omega_{wet}}{2\pi}\int_0^{\frac{\pi}{2}}{\left(\int_b^{a}{J_{n}(kr \rm{sin}\xi)\chi(r)rdr}\right)^{2}\rm{sin}\xi d\xi}
\\ \times \int_0^{2\pi}{{\rm{cos}}^{2}(n\psi){\rm{d}}{\psi}}\int_0^{\frac{\omega_{wet}}{2\pi}}\frac{\partial}{\partial t}{\left({\rm{Re}}(i^{n}e^{ik(c_{s}t-R)})\right)}
\\\times\frac{\partial}{\partial R}{\left(\frac{{\rm{Re}}(i^{n}e^{ik(c_{s}t-R)})}{R}\right)}{\rm{dt}}.
\end{multline}
The total energy pumped into the system by the source is
the mean kinetic energy of the plate and the adjacent fluid, \cite{Lamb,Amabiliannular}
\begin{equation}
E_{\rm{input}} = T_{P}(1+\beta_{mn})=V A^{2}(t),
\label{68}
\end{equation}
where $T_{P}$ is the mean kinetic energy of the resonator, and $\beta_{mn}$ is the added virtual mass incremental factor \cite{Amabiliannular} as mentioned in appendix \ref{A3}. The mean kinetic energy $T_{P}$ is given by,
\begin{multline}
T_{P} = \frac{1}{2}{\rho_{s}d_{0}\int_0^{2\pi}\int_b^a{({\chi(r)A(t)
\cos{n\theta}}})^2}r {\rm{d}}r {\rm{d}}\theta \\
= \frac{V}{2(1+\beta_{mn})}A^{2}(t).
\label{67}
\end{multline}
On equating the rate of decay of input energy from the source, i.e.,
\begin{equation}
\frac{dE_{\rm{input}}}{dt} = V A(t)\frac{dA(t)}{dt},
\label{69}
\end{equation}
with the rate of energy emitted into acoustic
field, i.e., eqn.~(\ref{66}), we get \cite{Lamb}.
\begin{equation}
\frac{dA(t)}{dt} = \delta A(t), \label{70}
\end{equation}
where $\delta=U/V$, the closed form expression for $\delta$ is given by,
\begin{multline}
\delta = -\frac{\rho_{f}}{\rho_{s}}\frac{R}{d_{0}}\frac{f_{wet}}{(1+\beta_{mn})}\frac{{\int_0^{\frac{\pi}{2}}{\left(\int_b^{a}{J_{n}(kr \rm{sin}\xi)\chi({\emph{r}}){\emph{r}}d{\emph{r}}}\right)^{2}\rm{sin}\xi d\xi}}}{\int_b^{a}\chi^{2}(r)r{\rm{d}}r} \\
\times \int_0^{\frac{\omega_{wet}}{2\pi}}\frac{\partial}{\partial t}{\left({\rm{Re}}(i^{n}e^{ik(c_{s}t-R)})\right)}\frac{\partial}{\partial R}{\left(\frac{{\rm{Re}}(i^{n}e^{ik(c_{s}t-R)})}{R}\right)}{\rm{d}}{\emph{t}}.
\end{multline}
Equation (\ref{70}) shows that the amplitude diminishes exponentially with an exponential constant $\delta$.
On equating this constant $\delta$ with the decay constant, $\zeta
\omega_{\rm{wet}}$ of the damped linear oscillator \cite{Rao}, we get $\zeta=\delta / \omega_{\rm{wet}}$. Based on the definition of quality factor \cite{Rao}, we get
\begin{equation}
\rm{Q}_{\rm{ac}} = \frac{1}{2\zeta} = \frac{\pi \emph{f}_{\rm{wet}}}{\delta} = \frac{\pi \emph{f}_{\rm{dry}}}{\delta \sqrt{1+\beta_{mn}}},
\label{71}
\end{equation}
where $\omega_{\rm{wet}}=2\pi f_{\rm{wet}}$, $f_{\rm{wet}}$ is the frequency of the vibrating plate in contact with the fluid. Although the final expression of the quality factor looks very simple, it requires the computations of $\delta$ and $\beta_{mn}$ which are mode dependent. Since the computation of these parameters requires successive differentiation and numerical integration, we perform these steps in MATLAB and MAPLE. The entire set of calculations is somewhat involved but algorithmic. We, therefore, present the sequence of calculations to be done as a schematic flow chart in Fig.~\ref{flowchart}.

There are three major computational blocks that we have tried to present separately for conceptual clarity. As is evident from Fig.~\ref{flowchart}, the computation involves one FEM analysis along with computer algebra and general numerics.

We can also use the procedure described above to compute the quality factor for a solid circular plate by setting $b=0$ and $\chi(r) = A_{mn}J_n(\beta r)+ C_{mn}I_n(\beta r)$. In the subsequent section, we compare the quality factor obtained using exact mode shapes as described in this paper with that using approximate mode shapes based on Lamb's formulation \cite{Lamb} for the first two modes of plate. Next, we use the analysis presented here to compute the quality factor for the annular resonator in higher modes, covering 15 modes. We also compare the computed quality factor with experimentally obtained values\cite{Darren} in the higher modes where acoustic radiation dominates the losses.

\section{\label{sec:level2}Results and Discussion}
We first compute the quality factor for a clamped solid circular plate vibrating in contact with the surrounding medium.
We concentrate on the first two modes. We compare our results with those of Lamb.

To do the analysis, we take the following dimensions and material properties of the structural and the fluid domains based on the test structure used by Southworth \emph{et al}. \cite{Darren} in their experimental studies. Here, the annular plate is made of silicon with Young's modulus $E=150$ GPa, the density $\rho_{s}=2330$ kg$/\rm{m}^{3}$, and the Poisson's ratio $\nu=0.22$. It has an outer radius of $a=18.4~\mu$m, inner radius of $b=2~\mu$m and a thickness of $d_{0}=300$ nm. The plate is surrounded by air having density $\rho_{f}=1.2$ kg$/\rm{m}^{3}$ under constant ambient temperature $T=293$ K and pressure $P=1.013 \times 10^{5}$ Pa. For the analysis of the solid plate, we simply take the inner radius to be zero, i.e., $b$ = 0 and rest of the parameters remain the same.

\subsection{\label{lambC} Solid circular plate}
\begin{figure}[h]
\includegraphics[width=9cm,clip]{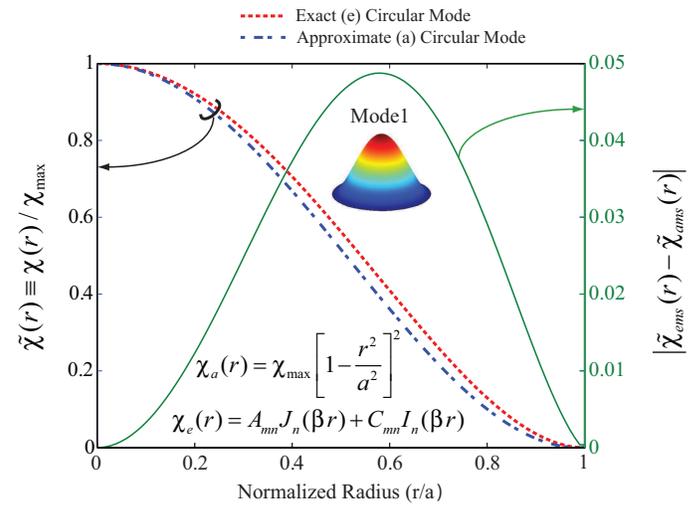}
\caption{(color online) Normalized circular mode shapes of a clamped solid circular plate}
\label{modeshapes1}
\end{figure}

\begin{figure}[h]
\includegraphics[width=9cm,clip]{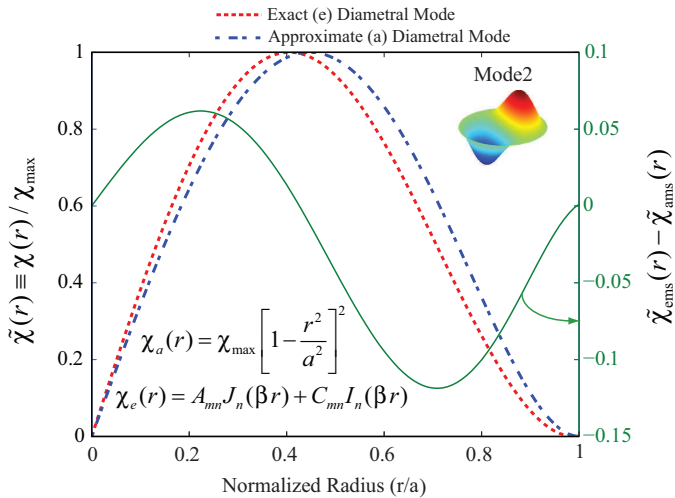}
\caption{(color online) Normalized diametral mode shapes of a clamped solid circular plate}
\label{modeshapes2}
\end{figure}

We compare the frequency and quality factor based on the exact mode shapes given by eqn.~(\ref{32}) and the approximate mode shapes used by Lamb for the first two modes. Fig.~\ref{modeshapes1} and \ref{modeshapes2} show the comparison between the normalized exact mode shape (ems) and the normalized approximate mode shape (ams) for the axisymmetric mode shape with zero nodal circle, i.e., $(m = 0, n = 0)$ and single nodal diameter mode $(m = 0, n = 1)$, respectively.
As is evident from these figures, the exact mode shape differs from the approximate one in spatial amplitude variation. The variation is significant enough to cause non-negligible differences in the radiated acoustic power. This is precisely what we see in the computed values of corresponding Q, listed in Table \ref{emsamsqcomparision}. The difference in amplitude variation across the plate between the two mode shapes causes very significant change in the corresponding Q. In particular, the approximate mode shape used by Lamb in the second mode overestimates the radiative losses by as much as 42\%. This approximation may lead to even bigger difference when we try to compute the Q's for an annular plate in the next section.
\begin{table}[h!]
\caption{\label{emsamsqcomparision}Quality factors associated with exact and approximate mode shapes corresponding to circular and diametral modes}
\begin{ruledtabular}
\begin{tabular}{cccccccc}
 $m$&$n$&$f_{\rm{wet}}$&$\rm{Q}_{\rm{ems}}$&$\rm{Q}_{\rm{ams}}$&$|\frac{\rm{Q}_{ems}-\rm{Q}_{ams}}{\rm{Q}_{ems}}|\%$\\ $$&$$&$\rm{(MHz)}$&$$&$$\\
 \hline
0   &0    &3.38  &116 &100 &13     \\
0   &1    &7.08  &233 &135 &42     \\
\end{tabular}
\end{ruledtabular}
\end{table}
\begin{table*}[hbtp]
\caption{\label{acousexptable}Experimental and theoretical comparison of acoustic Q-factors for annular
resonator clamped at the outer edge(\emph{a}) and free at the inner edge(\emph{b})
with \emph{b/a} = 0.1087}

\begin{ruledtabular}
\begin{tabular}{lccclcccccccc}
 \multicolumn{1}{c}{Mode No.}&\multicolumn{5}{c}{$A_{mn}J_n(\beta r)+B_{mn}Y_{n}(\beta r)+ C_{mn}I_n(\beta r)+D_{mn}K_{n}(\beta r)$}
 &\multicolumn{1}{c}{AVMI}&\multicolumn{1}{c}{$f_{wet}$}&\multicolumn{1}{c}{$f_{exp}$}&\multicolumn{2}{c}{Quality Factor}&$\%~\rm{Share}$\\
 $(m,n)$&$\beta$&$A_{mn}$&$B_{mn}$&$C_{mn}$&$D_{mn}$
&$\beta_{mn}$&$(\rm{MHz})$&$(\rm{MHz})$&$\rm{Q}_{\rm{ac}}$&$\rm{Q}_{\rm{exp}}$&$\rm{ of}\rm{~Q}_{\rm{ac}}$\\ \hline
   1(0,1)  &2.50149e5  &3.84177   &-0.03135  &0.05894     &0.03846      &0.09407   &7.06    &5.53    &236.42    &16.84   &7 \\
   2(0,2)  &3.18878e5  &-4.43796  &0.15639   &-0.02419    &0.10691      &0.00656   &11.49   &9.79    &260.02    &44.22   &17 \\
   3(0,3)  &3.87993e5  &-5.02600  &0.01979   &-0.01016    &0.01377      &0.00515   &17.02   &14.95    &206.13    &74.61   &36 \\
   4(1,1)  &4.20944e5  &4.89110   &-0.26138  &-0.00314    &0.40224      &0.00334   &20.05   &21.08    &214.78    &95.38   &44\\  \hline
   5(1,2)  &4.95634e5  &-5.37671  &0.40439   &0.00102     &0.27212      &0.00415   &27.79   &28.05    &161.15    &138.32  &85 \\
   6(1,3)  &5.71687e5  &-5.87366  &0.09854   &3.38038e-4  &0.07402      &0.00346   &36.98   &32.97    &208.94    &176.67  &85 \\
   7(0,7) &6.43180e5  &-6.90947  &1.35592e-7  &-5.03063e-4 &9.69105e-8   &0.00287   &46.83   &44.42    &340.89    &289.01  &85\\
   8(0,8) &7.04940e5  &-7.32436  &5.07846e-9  &-2.49051e-4 &3.65309e-9   &0.00259   &56.26   &53.84    &440.57    &264.04  &60\\
   9(0,9) &7.66200e5  &-7.72425  &1.79164e-10 &-1.24756e-4 &1.29718e-10  &0.00237   &66.47   &65.34   &584.40    &482.36  &82\\
   10(1,6) &7.80170e5  &-7.09124  &6.40735e-5 &1.97283e-5  &4.94654e-5   &0.00238   &68.92   &71.52    &584.33    &529.65  &90\\
   11(3,2)  &8.37278e5  &6.82720   &-1.18614  &-2.25503e-6 &0.08218      &0.00235   &79.38   &77.75    &703.59    &581.81  &82\\
   12(2,6)  &9.66106e5  &-7.72737  &5.33342e-4 &-5.60824e-7 &4.52572e-4   &0.00192   &105.71  &105.10   &984.80   &782.46  &79\\
  13(0,14) &10.67211e5  &9.55483   &2.51932e-14 &4.47644e-6  &1.77416e-14  &0.00167   &129.01  &124.64  &1101.62  &830.39  &75\\
   14(3,6) &11.46563e5 &-8.32963 &0.00282 &1.87631e-8 &0.00265  &0.00161 &148.91 &143.28 &1202.54 &966.97 &80 \\
   15(3,7) &12.18265e5  &-8.62869  &2.67745e-4  &6.57974e-9  &2.52229e-4   &0.00150   &168.13  &162.58   &1654.50 &1244.89 &75\\
\end{tabular}
\end{ruledtabular}
\newpage
\end{table*}
\subsection{\label{lambC} Annular plate}
We now consider the annular plate shown in Fig.~\ref{Acous}. We intend to compare our analytically computed results with experimental results of Southworth \emph{et al.} \cite{Darren}. Therefore, we take the geometry and the material properties used in Southworth \emph{et al.} for their drum resonator. For this resonator, the annular plate has fixed outer edge with the ratio of inner to outer radii as $0.1$. Moreover, the lower surface of the plate is close to another fixed-fixed plate but the upper surface is open to the surrounding air. Based on this geometric configuration, there are two major sources of energy dissipation: the squeeze film damping due to the trapped air film in the cavity below the resonator and the acoustic damping due to the acoustic radiation in the free space above the resonator surface.
We use the procedure outlined in section II to obtain the exact mode shapes of the resonator and use these mode shapes to compute the corresponding Q for each mode. The results are shown in Fig.~\ref{acousexptable}. We also include the experimental results reported by Southworth \emph{et al.} in this table. Our focus here is on computing  $\rm{Q}_{\rm{ac}}$ - the quality factor associated with acoustic radiation losses - and understanding its variation in different modes of vibration. Because acoustic radiation is not the only damping mechanism, we do not expect to match the experimentally obtained Q values with the computed $\rm{Q}_{\rm{ac}}$ values here. A comparison of the two Q's listed in Fig.~\ref{acousexptable} clearly indicates that the relative contribution of $\rm{Q}_{\rm{ac}}$ is very low in the lower modes of vibration and becomes dominant in the higher modes. It is well known that squeeze film damping is high at lower frequencies \cite{Pratap} and hence probably contributes maximum to the damping up to the $\rm{4}^{th}$ mode of vibration.
\begin{figure}[h]
\includegraphics[width=8cm,clip]{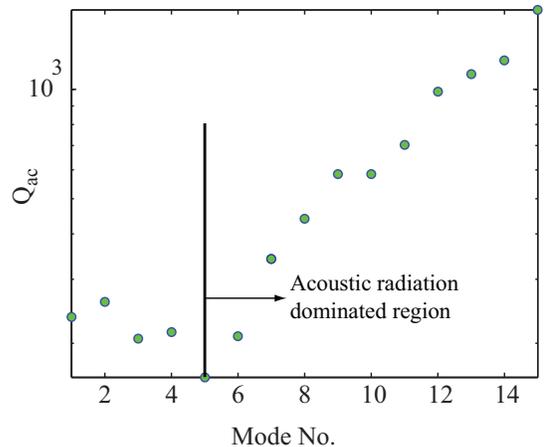}
\caption{(color online) Computed acoustic quality factors for various modes}
\label{Qvsmodenumber}
\end{figure}
Fig.~\ref{Qvsmodenumber} shows the computed values of $\rm{Q}_{\rm{ac}}$ graphically. It is interesting to see that the variation in $\rm{Q}_{\rm{ac}}$ is not monotonic with increasing frequency over the entire frequency range. It has a markedly different behavior in the lower modes (upto $\rm{5}^{th}$ mode), but settles down to a more predictable, almost monotonic trend, in the higher modes. The unsettled behavior in the lower modes can be explained qualitatively by looking at the corresponding mode shapes in Fig.~\ref{acousexptable} and pondering over the relative efficiency of these modes in transferring the resonator's energy into the acoustic radiative field. As the mode number increases, the resonator starts behaving like many point sources and addition of more such sources with increasing mode number makes less and less difference to $\rm{Q}_{\rm{ac}}$ leading to a somewhat predictable, monotonic and slower increase in $\rm{Q}_{\rm{ac}}$.

We know that $1/\rm{Q}_{\rm{net}} = 1/\rm{Q}_{\rm{ac}}+1/\rm{Q}_{\rm{rest}}$. We use this relationship to evaluate the relative contribution of $\rm{Q}_{\rm{ac}}$ to $\rm{Q}_{\rm{net}}$ (which is the $\rm{Q}_{\rm{exp}}$ here). The last column of Fig.~\ref{acousexptable} lists this contribution as a percentage of $\rm{Q}_{\rm{net}}$(computed from $\rm{Q}_{\rm{exp}}/\rm{Q}_{\rm{ac}}$ because of reciprocal realtionship). This column clearly shows the dominance of $\rm{Q}_{\rm{ac}}$ in higher modes.
\begin{figure}[h]
\includegraphics[width=8cm,clip]{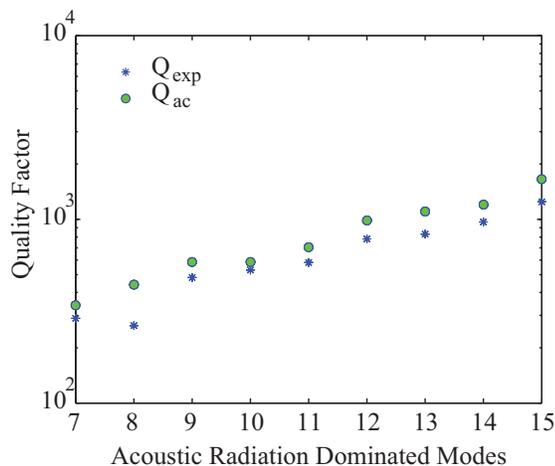}
\caption{(color online) Comparison of experimental and computed quality factors}
\label{Qexpandcompvsmodenumber7plus}
\end{figure}
Fig.~\ref{Qexpandcompvsmodenumber7plus} is a graphical representation of $\rm{Q}_{\rm{ac}}$ and $\rm{Q}_{\rm{exp}}$ in the acoustic radiation dominated modes. This figure shows that $\rm{Q}_{\rm{ac}}$ and $\rm{Q}_{\rm{exp}}$ are close to each other in these modes and thus the net Q is predominantly dependent on the acoustic radiation. Hence any effort to enhance Q in these modes should focus on decreasing acoustic radiation losses.

\begin{figure}[h]
\includegraphics[width=8cm,clip]{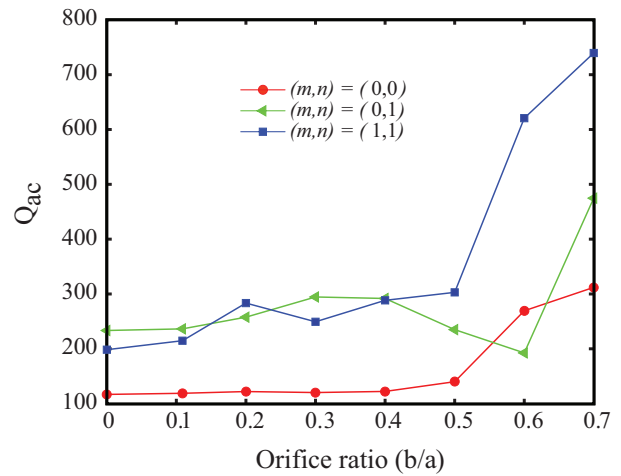}
\caption{(color online) Effect of the central orifice size on the acoustic radiation damping in a few selected modes of vibration. The modes chosen here are $(m,n)$ = (0,0), (0,1) and (1,1), taken purely as representative samples for circular and diametral modes.}
\label{Qacvsbbyadiffmn}
\end{figure}

 We also investigate the influence of the ratio of inner and outer radii on the quality factor. We observe a marginal increase in the acoustic quality factor with increasing radii ratio for all representative resonant modes shown in Table \ref{orificesizeacousticmode}. It is not until $b/a = 0.7$ (representing about 1/2 of the plate area as the hole) that we see a significant change in $\rm{Q}_{\rm{ac}}$, particularly in lower resonant modes (see Fig.~\ref{Qacvsbbyadiffmn}). At higher modes, the increase in $\rm{Q}_{\rm{ac}}$ is marginal even for this large $b/a$ ratio. This is along expected lines as the resonator radiates less and less acoustic power in the higher modes and any reduction in the resonator area results in much smaller reduction in the effective radiating area.
\begin{table}
\caption{\label{orificesizeacousticmode}Effect of orifice size and modal parameters ($m,n$) on the acoustic damping of the resonator}
\begin{ruledtabular}
\begin{tabular}{llccccc}
&&\multicolumn{5}{c}{\emph{$\rm{Q}_{ac}$ \rm{for various} \emph{b/a}}}\\
 $m$&$n$&$0.0$&$0.1087$&$0.3$&$0.5$&$0.7$\\ \hline
0   &0    &116.83  &119.01   &120.33 &140.11 &311.99     \\
0   &1    &233.41  &236.42   &294.70 &235.13 &474.85     \\
0   &2    &247.54 &260.02   &322.05 &320.74 &511.47    \\
0   &3    &205.10  &206.13   &243.33 &302.45 &297.01     \\
1   &1    &198.43 &214.78    &249.30   &303.04 &739.40      \\
1   &2    &157.47  &161.15   &199.10 &354.23 &882.70   \\
\end{tabular}
\end{ruledtabular}
\end{table}
\section{\label{sec:level1}Conclusions:\protect}
In this paper, we have presented a procedure for computing acoustic damping in various modes of vibration of an annular plate with fixed outer edge using exact mode shapes. A solid circular plate is treated as just a special case with the inner radius set to zero. The results for a solid circular plate are compared with those of Lamb where he used approximate mode shapes and gave expressions for acoustic radiation losses in the first two modes. We have shown that the difference in the exact mode shape and the approximate mode shape results in considerable difference in the Q-factor, particularly in the second mode. We have also computed the Q-factor associated with acoustic radiation for various modes of a micromechanical drum resonator and compared the results with published experimental values where the experimental results report the net Q in various modes. We have shown that the contribution of $\rm{Q}_{\rm{ac}}$ to the net Q is dominant for higher modes (above the \rm{5}$^{\rm{th}}$ mode of vibration) and accounts for almost 80\% of the net Q. This result shows that for two dimensional micromechanical resonators, one should use those modes of vibration that emit least amount of acoustic energy to obtain highest possible Q.

\section*{\label{sec:level1}Acknowledgements:\protect}
We thank Prof. Harold Craighead, D. R. Southworth and Robert Anthony Barton from Center for Materials Research, Cornell University for sharing the experimental results. We thank Prof. V. Sonti from Department of  Mechanical Engineering, Indian Institute of Science for useful discussion on Technical Acoustics. Research at Cornell was supported by the NSF through ECCS 0H1001742.

\appendix
\section{Solution procedure to find $w(r,\theta)$ \label{A00}}
The modal solution $w(r,\theta)$ of the eqn.~(\ref{6}) can be obtained by superimposing the solutions $w_{1}(r,\theta)$ and $w_{2}(r,\theta)$ of the equations,
\begin{equation}
 \nabla^{2}(w_{i}(r,\theta))\pm\beta^{2} w_{i}(r,\theta) = 0, \label{8}
\end{equation}
where, $i=1,2$. Assuming the solution of the form ${w}_{i}(r,\theta) =
{\chi}_{i}(r).{\psi}_{i}(\theta)$ based on the separable variable method and substituting it in eqn.~(\ref{8}), we obtain,
\begin{equation}
\frac{r^{2}}{\chi_{i}(r)}\left[\frac{ d^{2} \chi_{i}(r)}{d r^{2}}+\frac{1}{r}\frac{d \chi_{i}(r)}{d r}\pm \beta^{2}\chi_{i}(r)\right]
+\frac{1}{\psi_{i} (\theta)}\frac{d^2 \psi_{i}(\theta)}{d \theta^{2}} = 0
\label{10}.
\end{equation}
Due to the symmetric nature of the problem \cite{JRao}, periodicity in the azimuthal angle is introduced by equating the
second term in the above expression to the separation constant $-n^{2}$, we get,
\begin{multline}
\frac{1}{\psi_{i} (\theta)}\frac{d^2 \psi_{i}(\theta)}{d
\theta^{2}} = -n^{2} = \\
- \frac{r^{2}}{\chi_{i}(r)}\left[\frac{ d^{2} \chi_{i}(r)}{d
r^{2}}+\frac{1}{r}\frac{d \chi_{i}(r)}{d r}\pm
\beta^{2}\chi_{i}(r)\right] \label{11}.
\end{multline}
Under this condition, the modal solution can be written as $w(r,\theta) =
(\chi_{1}(r)+\chi_{2}(r))\psi(\theta)$. The equation~(\ref{11})
is decomposed into two Bessel's equations \cite{Watson} in $\chi_{i}$ and the simple
harmonic equation\cite{JRao} in $\psi(\theta)$ as follow,
\begin{equation}
\frac{ d^{2} \chi_{i}(r)}{d r^{2}}+\frac{1}{r}\frac{d \chi_{i}(r)}{d r}\pm \left(\beta^{2}\mp\frac{n^{2}}{r^{2}}\right)\chi_{i}(r) = 0,
\label{12}
\end{equation}
and
\begin{equation}
\frac{d^2
\psi(\theta)}{d \theta^{2}} + n^{2}\psi(\theta) = 0,
\label{13}
\end{equation}
where, $n$ is the number of nodal diameters and it takes a value of only positive integer, i.e., $n = 0,1,2,...$. The solution of equation (\ref{13}) is $\psi(\theta) =
C \cos{(n\theta+\epsilon)}$, where $A$ and $\epsilon$ are arbitrary
constants (for simplicity $\epsilon = 0$). The two Bessel's equations,
\begin{equation}\frac{ d^{2} \chi_{1}(r)}{d
r^{2}}+\frac{1}{r}\frac{d \chi_{1}(r)}{d r}+
\left(\beta^{2}-\frac{n^{2}}{r^{2}}\right)\chi_{1}(r) = 0,
\label{14}
\end{equation}
and
\begin{equation}
\frac{ d^{2}
\chi_{2}(r)}{d r^{2}}+\frac{1}{r}\frac{d
\chi_{2}(r)}{d r}-
\left(\beta^{2}+\frac{n^{2}}{r^{2}}\right)\chi_{2}(r) = 0,
\label{15}
\end{equation}
have the solutions of the form \cite{Watson},
\begin{equation}
\chi_{1}(r) = A_{mn}J_{n}(\beta r)+B_{mn}Y_{n}(\beta r),
\label{16}
\end{equation}
and
\begin{equation}
\chi_{2}(r) = C_{mn}I_{n}(\beta r)+D_{mn}K_{n}(\beta r),
\label{17}
\end{equation}
where $J_{n}$ is the Bessel function of the first kind, $Y_{n}$ is the Bessel function of the second kind and $A_{mn}$ and $B_{mn}$ are arbitrary constants in the solution of $\chi_{1}(r)$. $I_{n}$ and $K_{n}$ are the modified Bessel's function of first and second kind, respectively, and $C_{mn}$ and $D_{mn}$ are arbitrary constants in the solution of $\chi_{2}(r)$. Consequently, the final form of the modal solution $w(r,\theta)$ is found as, \cite{Watson}
\begin{multline}
w(r,\theta) = [A_{mn}J_n(\beta r)+B_{mn}Y_{n}(\beta r)+\\
C_{mn}I_n(\beta r)+D_{mn}K_{n}(\beta r)]\psi(\theta) = \chi(r)\psi(\theta). \label{18}
\end{multline}

\section{System of linear and homogenous equations for the four constants $A_{mn}$, $B_{mn}$, $C_{mn}$ and $D_{mn}$ \label{A0}}

\begin{equation}
A_{mn}J_{n}(\beta
a)+B_{mn}Y_{n}(\beta a)+C_{mn}I_{n}(\beta
a)+D_{mn}K_{n}(\beta a) = 0,
\label{19}
\end{equation}
\begin{multline}
A_{mn}\left[\frac{n}{\beta a}J_n(\beta a)-J_{(n+1)}(\beta
a)\right]+\\B_{mn}\left[\frac{n}{\beta a}Y_n(\beta
a)-Y_{(n+1)}(\beta a)\right] +\\C_{mn}\left[\frac{n}{\beta
a}I_n(\beta a)+I_{(n+1)}(\beta
a)\right]+\\D_{mn}\left[\frac{n}{\beta a}K_n(\beta
a)-K_{(n+1)}(\beta a)\right] \\ = 0 \label{20},
\end{multline}
%
%
\begin{multline}
A_{mn}F_{1}(\nu,n,\beta b)+B_{mn}F_{2}(\nu,n,\beta b) \\
-C_{mn}F_{3}(\nu,n,\beta b)-D_{mn}F_{4}(\nu,n,\beta b) = 0,
\label{21}
\end{multline}
%
\begin{multline}
A_{mn}\phi_{1}(\nu,n,\beta b)+B_{mn}\phi_{2}(\nu,n,\beta b) \\
-C_{mn}\phi_{3}(\nu,n,\beta b)-D_{mn}\phi_{4}(\nu,n,\beta b) = 0,
\label{22}
\end{multline}
where, the functions $F_{1},F_{2},F_{3},F_{4},\phi_{1},\phi_{2},\phi_{3}
$ and $\phi_{4}$ are defined by eqns.~(\ref{23})-(\ref{30}) as mentioned on the appendix~\ref{A1}.
\section{Expressions for functions $F_{1},F_{2},F_{3},F_{4},\phi_{1},\phi_{2},\phi_{3}
$ and $\phi_{4}$ \label{A1}}
\begin{multline}
F_{1}(\nu,n,\beta b) = \\
\left(J_{n}(\beta b)-(1-\nu)\left[\frac{n(n-1)}{(\beta
b)^{2}}J_{n}(\beta b) +\frac{1}{\beta b}J_{(n+1)}(\beta b)\right]
\right) \label{23}
\end{multline}

\begin{multline}
F_{2}(\nu,n,\beta b) = \\
\left(Y_{n}(\beta b)-(1-\nu)\left[\frac{n(n-1)}{(\beta
b)^{2}}Y_{n}(\beta b) +\frac{1}{\beta b}Y_{(n+1)}(\beta b)\right]
\right) \label{24}
\end{multline}

\begin{multline}
F_{3}(\nu,n,\beta b) = \\
\left(I_{n}(\beta b)+(1-\nu)\left[\frac{n(n-1)}{(\beta
b)^{2}}I_{n}(\beta b) -\frac{1}{\beta b}I_{(n+1)}(\beta b)\right]
\right) \label{25}
\end{multline}

\begin{multline}
F_{4}(\nu,n,\beta b) = \\
 \left(K_{n}(\beta
b)+(1-\nu)\left[\frac{n(n-1)}{(\beta b)^{2}}K_{n}(\beta b)
+\frac{1}{\beta b}K_{(n+1)}(\beta b)\right] \right) \label{26}
\end{multline}

\begin{multline}
\phi_{1}(\nu,n,\beta b) = \\
nJ_n(\beta b)-(\beta b)J_{(n+1)}(\beta
b)\\
+\frac{n^{2}(1-\nu)}{(\beta b)^{2}}\left[(n-1)J_{n}(\beta
b)-(\beta b)J_{(n+1)}(\beta b)\right] \label{27}
\end{multline}
\begin{multline}
\phi_{2}(\nu,n,\beta b) = nY_n(\beta b)-(\beta b)Y_{(n+1)}(\beta
b)\\
+\frac{n^{2}(1-\nu)}{(\beta b)^{2}}\left[(n-1)Y_{n}(\beta
b)-(\beta b)Y_{(n+1)}(\beta b)\right] \label{28}
\end{multline}
\begin{multline}
\phi_{3}(\nu,n,\beta b) = nI_n(\beta b)+(\beta b)I_{(n+1)}(\beta
b)\\
-\frac{n^{2}(1-\nu)}{(\beta b)^{2}}\left[(n-1)I_{n}(\beta
b)+(\beta b)I_{(n+1)}(\beta b)\right] \label{29}
\end{multline}
\begin{multline}
\phi_{4}(\nu,n,\beta b) = nK_n(\beta b)-(\beta b)K_{(n+1)}(\beta
b)\\
-\frac{n^{2}(1-\nu)}{(\beta b)^{2}}\left[(n-1)K_{n}(\beta
b)-(\beta b)K_{(n+1)}(\beta b)\right] \label{30}
\end{multline}
\section{Determinant of the coefficient matrix of eqns.~{\ref{19}}-{\ref{22}}\label{A2}}
\begin{widetext}
\begin{align}
\hspace{-25 mm}\left| \begin{array}{cccc}
J_n(\beta
a) & Y_{n}(\beta a) & I_{n}(\beta a) &K_{n}(\beta a)\\
\frac{n}{\beta a}J_n(\beta a)-J_{(n+1)}(\beta
a) &\frac{n}{\beta a}Y_n(\beta a)-Y_{(n+1)}(\beta
a) &\frac{n}{\beta a}I_n(\beta a)+I_{(n+1)}(\beta
a) &\frac{n}{\beta a}K_n(\beta a)-K_{(n+1)}(\beta
a) \\
F_{1}(\nu,n,\beta b)
& F_{2}(\nu,n,\beta b)
&-F_{3}(\nu,n,\beta b)
&-F_{4}(\nu,n,\beta b)\\
\phi_{1}(\nu,n,\beta b)
& \phi_{2}(\nu,n,\beta b)
&-\phi_{3}(\nu,n,\beta b)
&-\phi_{4}(\nu,n,\beta b)
\end{array} \right|= 0
\end{align}
\end{widetext}{\label{annu1}}

\section{Definition of $\beta_{mn}$ \label{A3}}

The frequency of free vibration of the plate in fluid can be related to its natural frequency in vacuum and is written as, \cite{Amabiliannular}
\begin{equation}
f_{wet} = \frac{f_{dry}}{\sqrt{1+\beta_{mn}}},
\label{43}
\end{equation}
where, $\beta_{mn}$ is called added virtual mass incremental(AVMI)
factor, \cite{Amabiliannular} as
\begin{equation}
\beta_{mn} = \frac{T_{F}}{T_{P}} = \Gamma_{mn}\frac{\rho_{f}}{\rho_{s}}\frac{a}{d_{0}}.
\label{44}
\end{equation}
$\beta_{mn}$ is defined as the ratio of reference kinetic energy
$T_{F}$ of the fluid, due to the structural vibration, and that of
the resonator $T_{P}$. $\Gamma_{mn}$ is the non-dimensional added
virtual mass incremental (NAVMI) factor. Note the above NAVMI
factor is for plate in contact with fluid only on one side
(Fig.~\ref{Acous}). NAVMI factors obtained must be doubled
for plates in contact with a fluid on both sides. NAVMI factors
are found to be given by,
\begin{equation}
\Gamma_{mn} = \frac{1}{\beta_{mn}}\frac{d_{0}}{a}\frac{\rho_{s}}{\rho_{f}} = \int_0^\infty H^{2}_{mn}(\eta) d\eta,
\label{45}
\end{equation}
where
\begin{multline} 
H_{mn}(\eta)=A_{mn}H_{Amn}(\eta)+B_{mn}H_{Bmn}(\eta)\\+C_{mn}H_{Cmn}(\eta)+D_{mn}H_{Dmn}(\eta),
\label{46}
\end{multline}
in which,
\begin{multline}
H_{Amn}(\eta) = \frac{1}{\lambda^{2}-{\eta^{2}}}[\lambda
J_{n}(\eta)J_{n+1}(\lambda)-\eta J_{n+1}(\eta)J_{n}(\lambda)]
\\-\frac{b}{a(\lambda^{2}-{\eta^{2}})}\left[\lambda J_{n}\left(\eta \frac{b}{a}\right)J_{n+1}\left(\lambda
\frac{b}{a}\right)\right]\\-\frac{b}{a(\lambda^{2}-{\eta^{2}})}\left[-\eta
J_{n+1}\left(\eta \frac{b}{a}\right)J_{n}\left(\lambda
\frac{b}{a}\right)\right], \label{47}
\end{multline}
\begin{multline}
H_{Bmn}(\eta) = \frac{1}{\lambda^{2}-{\eta^{2}}}[\lambda
J_{n}(\eta)Y_{n+1}(\lambda)-\eta J_{n+1}(\eta)Y_{n}(\lambda)]
\\-\frac{b}{a(\lambda^{2}-{\eta^{2}})}\left[\lambda J_{n}\left(\eta \frac{b}{a}\right)Y_{n+1}\left(\lambda
\frac{b}{a}\right)\right]\\-\frac{b}{a(\lambda^{2}-{\eta^{2}})}\left[-\eta
J_{n+1}\left(\eta \frac{b}{a}\right)Y_{n}\left(\lambda
\frac{b}{a}\right)\right], \label{48}
\end{multline}
\begin{multline}
H_{Cmn}(\eta) = \frac{1}{\lambda^{2}+{\eta^{2}}}[\lambda
J_{n}(\eta)I_{n+1}(\lambda)+\eta J_{n+1}(\eta)I_{n}(\lambda)]
\\-\frac{b}{a(\lambda^{2}+{\eta^{2}})}\left[\lambda J_{n}\left(\eta \frac{b}{a}\right)I_{n+1}\left(\lambda
\frac{b}{a}\right)\right]\\-\frac{b}{a(\lambda^{2}+{\eta^{2}})}\left[\eta
J_{n+1}\left(\eta \frac{b}{a}\right)I_{n}\left(\lambda
\frac{b}{a}\right)\right], \label{49}
\end{multline}
\begin{multline}
H_{Dmn}(\eta) = \frac{-1}{\lambda^{2}+{\eta^{2}}}[\lambda
J_{n}(\eta)K_{n+1}(\lambda)-\eta J_{n+1}(\eta)K_{n}(\lambda)]
\\-\frac{b}{a(\lambda^{2}+{\eta^{2}})}\left[-\lambda J_{n}\left(\eta \frac{b}{a}\right)K_{n+1}\left(\lambda
\frac{b}{a}\right)\right]\\-\frac{b}{a(\lambda^{2}+{\eta^{2}})}\left[\eta
J_{n+1}\left(\eta \frac{b}{a}\right)K_{n}\left(\lambda
\frac{b}{a}\right)\right]. \label{50}
\end{multline}
and the above stated equations are used for annular resonators.
For solid disk resonator, we have,
\begin{equation}
H_{mn}(\eta)=A_{mn}H_{Amn}(\eta)+C_{mn}H_{Cmn}(\eta),
\label{51}
\end{equation}
\begin{equation}
H_{Amn}(\eta) = \frac{1}{\lambda^{2}-{\eta^{2}}}[\eta
J_{n-1}(\eta)J_{n}(\lambda)\\-\lambda J_{n-1}(\lambda)J_{n}(\eta)],
\label{52}
\end{equation}
\begin{equation}
H_{Cmn}(\eta) = \frac{1}{\lambda^{2}+{\eta^{2}}}[\eta
J_{n+1}(\eta)I_{n}(\lambda)\\+\lambda J_{n}(\eta)I_{n+1}(\lambda)].
\label{53}
\end{equation}

The velocity potential '$\phi$' at the plate fluid interface is
obtained by using the above $H_{mn}$ function, which has
eliminatable singularity at $\eta = \lambda$ (Here $\lambda = \beta
a$)\cite{Amabilisolid,Amabiliannular},
\begin{equation}
\phi(\alpha,0) = -a\int_0^\infty{H_{mn}(\eta)J_{n}(\eta\alpha)d\eta}.
\label{54}
\end{equation}
The derivation of velocity potential assumes the fluid such as air
to be incompressible, inviscid newtonian fluid. The plate is
considered placed in an annular aperture of an infinite rigid
wall. The fluid motion, considered only due to the plate's
vibration, is assumed to be irrotational. The plate and fluid are
maintained at ambient isothermal conditions. Hence such velocity
potential must satisfy Laplace equation. By compatibility
conditions velocity of the fluid in contact with the plate assumes
the plate velocity. No slip flows occur at the fluid-structure
interface. Such uncoupled fluid-structure interaction problem
involves evaluation of reference kinetic energy of the fluid at
the plate-fluid interface by using the above velocity potential
and hence this kinetic energy in-turn is used to evaluate
$\beta_{mn}$. The integral in equation(\ref{45}) was numerically
evaluated in MATLAB using the adaptive Gauss-Kronrod quadrature.

\end{document}